\documentclass[%
 aip,
 amsmath,amssymb,
 reprint%
]{revtex4-2}
\usepackage{gensymb}
\usepackage{graphicx}
\usepackage{dcolumn}
\usepackage{bm}
\usepackage{etoolbox}
\usepackage{xcolor}
\usepackage[colorlinks=true,linkcolor=blue,citecolor=red]{hyperref}%
\usepackage{siunitx}
\usepackage{subfig}
\usepackage{mathtools}
\usepackage{CJKutf8}
\newcommand*{\Cn}[1]{\begin{CJK}{UTF8}{gbsn}#1\end{CJK}}
\newcommand*{\Ja}[1]{\begin{CJK}{UTF8}{ipxm}#1\end{CJK}}

\newcommand{\figref}[1]{Fig.\,\ref{#1}}

\newcommand{\figrefp}[2]{Fig.\,\ref{#1}\,(#2)}

\makeatletter
\def\@email#1#2{%
 \endgroup
 \patchcmd{\titleblock@produce}
  {\frontmatter@RRAPformat}
  {\frontmatter@RRAPformat{\produce@RRAP{*#1\href{mailto:#2}{#2}}}\frontmatter@RRAPformat}
  {}{}
}%
\makeatother
\begin{document}

\preprint{AIP/123-QED}

\title[Flowing instability]{
Ridge instability in dense suspensions caused by the second normal stress difference}

\author{Zhongqiang Xiong (\Cn{熊钟强})}

\affiliation{Wenzhou Key Laboratory of Biomaterials and Engineering, Wenzhou Institute, University of Chinese Academy of Sciences, Wenzhou, 325000, China}

\author{Peter Angerman}%
\affiliation{Complex Fluids Research Group, Department of Chemical Engineering, Swansea University, Swansea SA1 8EN, United Kingdom}
\affiliation{Basque Center for Applied Mathematics (BCAM), Alameda de Mazarredo 14, 48009 Bilbao, Spain}%

\author{Marco Ellero}%
 
\affiliation{Basque Center for Applied Mathematics (BCAM), Alameda de Mazarredo 14, 48009 Bilbao, Spain}%
\affiliation{IKERBASQUE, Basque Foundation for Science, Calle de Maria Dias de Haro 3, Bilbao, 48013, Bizkaia, Spain}%
\affiliation{Zienkiewicz Center for Computational Engineering (ZCCE), Swansea University, Bay Campus, Swansea, SA1 8EN, United Kingdom}%

\author{Bjornar Sandnes}
\affiliation{Complex Fluids Research Group, Department of Chemical Engineering, Swansea University, Swansea SA1 8EN, United Kingdom}%

\author{Ryohei Seto (\Ja{瀬戸亮平})}%
\affiliation{Wenzhou Key Laboratory of Biomaterials and Engineering, Wenzhou Institute, University of Chinese Academy of Sciences, Wenzhou, 325000, China}
\affiliation{Oujiang Laboratory (Zhejiang Lab for Regenerative Medicine, Vision and Brain Health), Wenzhou, 325000, China}
\affiliation{Graduate School of Information Science, University of Hyogo, Kobe, 650-0047, Japan}
\email{seto@ucas.ac.cn}
\date{Submitted: 16 November 2023. Accepted: 14 January 2024. doi: 10.1063/5.0188004}

\begin{abstract}
A dense suspension of the cornstarch flowing on a very inclined wall finally forms some ridge-like patterns of the free surface. 
The onset of pattern formation is the primary target to elucidate the mechanism. 
In this work, based on the continuity of fluids and the force balance, we show that the flat free surface is unstable when the second normal stress difference $N_2$ is negatively proportional to shear stress and the gravity component perpendicular to the wall is weak enough. 
Such instability is inevitable to grow into a ridge-like surface profile oriented parallel to the flow direction.
%
%
We use the instability criterion to predict the critical slope angle for the formation of ridge patterns. 
The estimated critical angle was found to be in agreement with experimental observations for a cornstarch suspension.
\end{abstract}

\maketitle

\section{Introduction}
\label{sec: intro}

The thin liquid film flowing down on an inclined plane (windowpane, guttering, slope, etc.) driven by gravity is common in our daily lives.
The most fascinating phenomenon here is that some regular wave patterns or morphogenesis of the free surface appear. 
To understand the mechanism behind is related to the flowing instability indicated in the nonlinearity or the coupling of the governing equations, which is usually related to the fluid properties. 
The inertia, capillary, and surface-tension gradient are essential to such pattern formations for Newtonian fluids. 
A review paper by \citet{RevModPhys.81.1131} has addressed these points comprehensively. 
For non-Newtonian fluids, the morphogenesis is even more diverse and complex.


In suspensions, a surface wave called roll wave has been observed in experiments with an inclined angle\,\cite{balmforth2005roll, darbois2020surface}.
%
The flow instability generating the dynamic pattern is demonstrated to be different from the classical Kapitza instability by inertia\,\cite{kapitza1948wave}. 
Such instability is of purely rheological origin, i.e., caused by discontinuous shear thickening (DST), where the negative slope in the stress vs.~shear rate curve is the key to the criterion of the instability\,\cite{darbois2020surface}. 
%



Besides the DST, another property of dense suspensions is the presence of normal stress difference\,\cite{ness2022physics}. 
%
For general non-Newtonian fluids,
the normal stress differences are detected by normal force via the rheometer with parallel-plate and cone-and-plate geometries. 
It is well known that normal stress differences can distort the free surface.
The most famous example is the classical Weissenberg effect, i.e., the first normal stress difference ($N_1 \coloneqq \sigma_{11}-\sigma_{22}$, where the 1, 2, and 3 denote the direction of `flow,' `gradient,' and `vorticity,' respectively) forms a rod climbing shape of the free surface around a rotating cylinder\,\cite{tanner2000engineering}.
%


The second normal stress difference ($N_2 \coloneqq \sigma_{22}-\sigma_{33}$) is also reported to cause a curved steady profile of the free surface by \citet{tanner1970some} and \citet{sturges1975slow} about fifty years ago.
The relation for $N_2$ is even designed to estimate the normal stress functions by detecting the steady height profile with cameras.
Particle-scale surface deformation
observed with suspensions flowing on an inclined plane also implies 
the presence of normal stresses\,\cite{Timberlake_2005}.
%
%
\citet{10.1122/1.551083} found $N_1/(\beta\tau)=-0.15 \pm 0.05$ and $N_2/(\beta\tau) = -0.54 \pm 0.03$, where $\tau=\sigma_{12}$, and $\beta = 2.17\phi^3 e^{2.34\phi}$ with the solid volume fraction $\phi$. 
\citet{singh2003experimental} used parallel-plate and Couette geometries to measure $N_2/\tau$, which are slightly larger than the above results at $\phi=0.45$. 
%
\citet{dai2013viscometric} then show that the best fits, in the range $0.1\leq\phi\leq0.45$,
are $N_1/\tau=-0.8\phi^3$ and $N_2/\tau=-4.4\phi^3$, respectively.
However, some factors can lead to systematic errors in the measurement of normal stresses, such as the pressure transducers, nonflush pressure holes of apparatuses, and the drowned edge used in the parallel-plate measurements\,\cite{tanner2000engineering}. 
Normal stress differences for dilute or non-dense suspensions ($\phi \lesssim 0.2$) are difficult to measure since they are much smaller than the shear stress. 
Because there are some difficulties in the measurements of normal stress differences in experiments, using a different combination of test methods is necessary to guarantee the validity of experimental measurements. 
\citet{cwalina2014material} observed $N_1/\tau$ and $N_2/\tau$ to be functions of $\phi$ in experiments.
Such ratios were assumed by \citet{10.1122/1.551021} in their particle migration model.
Later, \citet{Timberlake_2005} used the model to satisfactorily explain experimental observations of the particle migration in fully-developed suspension flows on an inclined plane.
More details for the normal stress differences of non-colloidal suspension can be found in Tanner's review paper\,\cite{tanner2018aspects}. 
The importance of the second normal stress difference is also highlighted by \citet{maklad2021review}.
%


Complementally, obtaining the normal stress differences in simulations is more straightforward.
\citet{sierou2002rheology} used accelerated Stokesian Dynamics to study the normal stress differences and the particle pressure of a monodisperse non-Brownian suspension with hydrodynamic interactions and a weak interparticle force under simple shear flow. 
Recently, contact friction has been recognized to play a major role in dense suspensions\,\cite{PhysRevLett.111.218301}.
\citet{mari2014shear} and \citet{gallier2014rheology} showed that friction of particles notably increases the viscosity and $N_2$ but affects $N_1$ relative to the shear stress less significantly, which is closer to the experimental results\,\cite{gallier2014rheology}. 
The nearly proportional relation between normal stress differences and shear stress is also observed in simulations\,\cite{Romain2015Discontinuous, seto2018normal}. 
Data, including experiments and simulations, were collected from different references, which is shown in a recent review paper by \citet{guazzelli2018rheology}
(see also \figref{fig:f3}).

The coupling between the normal stress difference and shear stress would be important to understand the complex flow.
Therefore, inspired by Tanner's work\,\cite{tanner1970some}, $N_2$ is assumed to be a critical factor in understanding the ridge patterns in dense suspensions.
One natural question is to determine the evolution of the flow from the initial flat configuration to the steady configuration based on the rheological properties of suspensions so that we can know the details of flowing dynamics throughout.
However, the inclined flow field will be much more complicated as there will be more distortions of the free surface. 
It is worth mentioning that $N_2$ is well defined in a standard simple shear flow and not invariant under a change of coordinates.
For instance, in a flow with a highly distorted free surface, we would need $N_2$ in a local coordinate\cite{10.1122/1.4986840} 
rather than the global coordinate system in \figrefp{ridge_pattern_experiment}{a}.
Also, the particle migration may become significant for fully developed flows\,\cite{Timberlake_2005}.
This is why we focus on the very early stage of the pattern formation, where the free surface is nearly flat so that the local coordinate coincides with the global coordinate we used, and the particle migration is negligible.
In this special case, we are able to use the linear instability analysis, which is powerful in analyzing the pattern formations of 
free surfaces\,\cite{RevModPhys.81.1131} and even for complex bioconvection 
patterns\,\cite{Chakraborty_2018}.
Therefore, we do not determine fully-developed surface profiles in this work.
%


If we are able to reproduce the evolution of the flow near the initial flat configuration, we can answer two questions in a concise way: 1. Why can the cornstarch suspension distort from the initial flat configuration along the perpendicular direction of flow (see \figref{ridge_pattern_experiment})? 2. What is the critical angle of inclination to destabilize the initial flat configuration?
%


%
The pattern formation process starts from a distortion fluctuation of a flat surface, and then the free surface keeps distorting with some instabilities. 
Therefore, identifying the underlying instability is prior to explaining the pattern formation. 
In this work, we predict the critical angle ($\theta_{\mathrm{c}}$) to form the ridge flow based on $N_2$ by linear instability analysis.
In section \ref{sec: experiment}, we describe the experimental setup.
In section \ref{subsec: continuity}, the continuity equation of fluids with the free surface is introduced. 
In section \ref{subsec: force balance}, the force balance equation related to the free surface profile is determined by the lubrication theory. 
Section \ref{subsec: constitutive relation} addresses the linear relation between $N_2$ and shear stress, the constitutive relation used in this work.
In section \ref{sec:s3}, by combining the ingredients demonstrated in the previous section, the origin of the instability can be identified to obtain the $\theta_{\mathrm{c}}$.

\section{Experiment} \label{sec: experiment}

\begin{figure*}[t]
\centering
\includegraphics[width=0.80\textwidth]{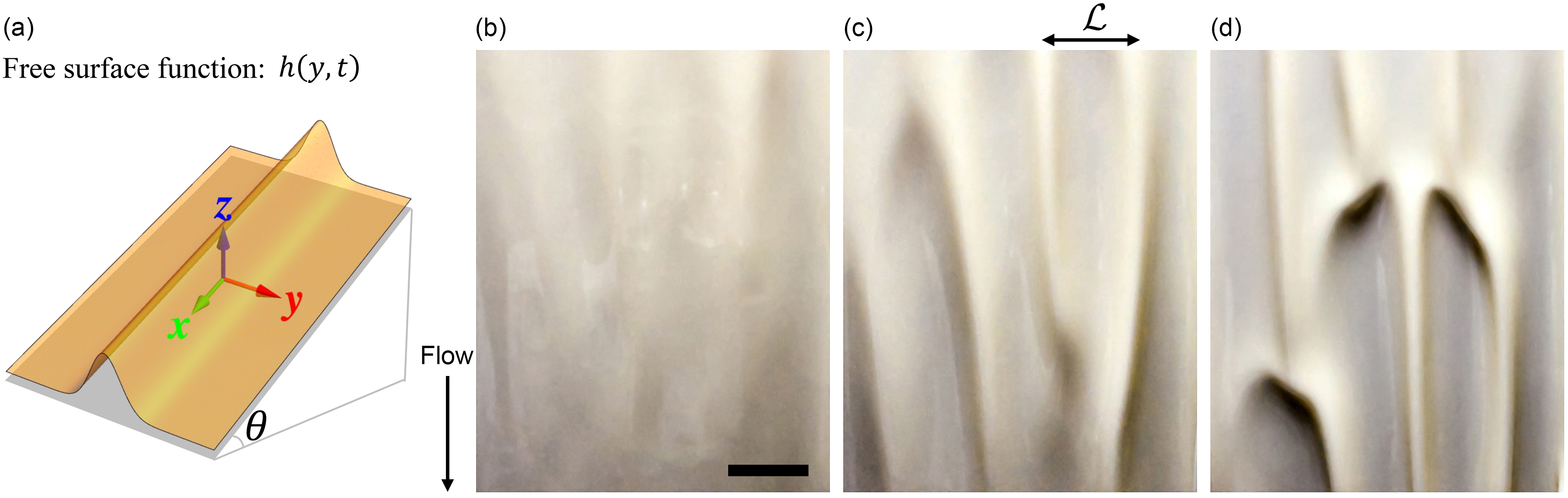}
\caption{\label{ridge_pattern_experiment}
(a) Sketch of ridge and coordinate system.
(b)--(d) Time series of ridge formation with \SI{10}{\second} duration between frames.
The flow was from top to bottom. 
$\theta = 87.5\degree$, $\phi \approx 0.37$, and the scale bar is \SI{5}{\cm}. 
The characteristic length between ridges is denoted $\mathcal{L}$, and the characteristic thickness scale $\mathcal{H}$ is the initial thickness of the film.}
\end{figure*}

Cornstarch suspension was prepared by mixing cornstarch particles with \SI{99.5}{\percent} glycerol.
The material volume fraction was calculated following the equation\,\cite{JaegerHanEndao}: 
\begin{equation}
\phi = 
\frac{(1-\beta)m_{\mathrm{cs}}/\rho_{\mathrm{cs}}}{(1-\beta)m_{\mathrm{cs}}/\rho_{\mathrm{cs}}+m_{\mathrm{l}}/\rho_{\mathrm{l}}
+\beta m_{\mathrm{cs}}/\rho_{\mathrm{w}}}
    \label{Jaeger VolFrac}
\end{equation}
where $\beta \approx 0.13$ accounts for the fraction of the measured `cornstarch' mass occupied by water due to humidity, $m_{\mathrm{cs}}$ is the measured mass of cornstarch, 
$\rho_{\mathrm{cs}} \approx \SI{1620}{\kilogram . \metre^{-3}}$ is the density of cornstarch, $m_{\mathrm{l}}$ is the measured mass of the solvent, $\rho_{\mathrm{l}} \approx \SI{1260}{\kilogram . \metre^{-3}}$ is the glycerol solvent density, 
and $\rho_{\mathrm{w}} \approx \SI{998}{\kilogram . \metre^{-3}}$ is the density of water. 
A sample with cornstarch mass fraction $x_{\mathrm{m}} \approx 0.50$ and the material volume fraction of $\phi \approx 0.37$ was prepared, which appears to be continuous shear thickening (CST), but not DST as expected (details in Appendix \ref{viscos}).

 
The sample was left to rest for 1 hour with intermittent mixing. 
Then, the suspension was transferred to an inclined plane setup consisting of a perspex tray mounted on a rotating axis. 
The suspension was allowed to settle in a uniform layer with an average film depth of approximately \SI{9}{\mm}. 
The dimensions of the flow plane were \SI{30}{cm} width and \SI{100}{cm} length.
Inclined flow experiments were carried out in increments of \ang{2.5}. 
The surface of the flowing film was filmed with a camera (Nikon 1 J4) mounted above the plane.
\figrefp{ridge_pattern_experiment}{a} illustrates the coordinate system and tilt angle. 
The experiment commenced by quickly tilting the plane to the desired angle. 
The undulation of the surface is visible by the degree of light reflection. 
%


At low $\theta$, the surface remained flat and unperturbed during the ensuing gravity-driven flow down the incline.
At high $\theta$, the surface was unstable, and ridges formed running parallel to gravity and the flow direction. 
Fig.\,\ref{ridge_pattern_experiment}\,(b)--(d) exhibit a time series of one experiment at $\theta = \ang{87.5}$, showing the gradual emergence and growth of the ridge pattern.

In our experiments with increasing $\theta$, the ridge pattern was first observed at $\theta = \ang{77.5}$, so we conclude that the critical angle is in the range $\ang{75.0} < \theta_{\mathrm{c}} < \ang{77.5}$ for this experiment.
%
%

\section{Problem Formulation} \label{sec:s2}

The ridge pattern in \figref{ridge_pattern_experiment} was obtained for $\phi \approx 0.37$,
which corresponds to the CST regime for the dense suspension. 
Note that no roll waves, which are a signature of the DST rheology requiring a higher $\phi$, were observed.
%
While cornstarch suspensions are known for their DST behavior, it is clear that the ridge instability is not caused by DST directly. 
We hypothesize that the instability is a consequence of the second normal stress difference $N_2$.


In this work, we focus on the instability along the $y$ direction, i.e., the free surface will only distort on the $y$-$z$ plane.
Therefore, we consider translational symmetry along the $x$ direction (see the illustration in \figrefp{ridge_pattern_experiment}{a}). 
As a consequence, the pressure scalar, velocity vector, and stress tensor of suspensions are functions of coordinate $y$ and $z$ only, 
i.e., $p=p(y,z,t)$, $\bm{v}=\bm{v}(y,z,t)$, and $\bm{\tau}=\bm{\tau}(y,z,t)$. 
The free surface configuration is defined by $z=h(y,t)$.

\subsection{Continuity equation of fluids with the free surface}
\label{subsec: continuity}

First of all, the law of conservation of mass for incompressible fluids is
\begin{equation}
\label{equ:mass_conservation}
    \nabla \cdot \bm{v}=0.
\end{equation}
Integrating both sides of eq.\,\eqref{equ:mass_conservation} in $z$, one has
\begin{align}
 0&=\int_0^h\frac{\partial v_y}{\partial y}dz+\int_0^h\frac{\partial v_z}{\partial z}dz \notag\\
 &=\frac{\partial}{\partial y}\int_0^hv_ydz-v_y|_{z=h}\frac{\partial h}{\partial y}+v_z|_{z=h},
 \label{equ:integ_v}
\end{align}
where the non-slip boundary condition $v_z|_{z=0}=0$ is used since the fluids contact the wall is stationary if there is no slip of the fluids.
On the other hand, the material points move and deform on the free surface. 
On the free surface, one has
\begin{equation}
 0=\frac{D}{Dt}\left(z-h(y,t)\right)=v_z|_{z=h}-\left(\frac{\partial h}{\partial t}+v_y|_{z=h}\frac{\partial h}{\partial y}\right),
 \label{equ:free_surface}
\end{equation}
where $D/Dt$ is the material derivative. 
Combined eqs.~\eqref{equ:integ_v} and \eqref{equ:free_surface}, therefore, the continuity equation of the free surface is
\begin{equation}\label{equ:free_surface_continuity}
\frac{\partial h}{\partial t}+\frac{\partial}{\partial y}\int_0^hv_ydz=0.
\end{equation}

\subsection{Force balance equation for thin fluid films}
\label{subsec: force balance}

The law of conservation of momentum, known as the equation of motion, is
\begin{equation}
\nabla \cdot \bm{\sigma}+\rho \bm{g}=
\rho \frac{D \bm{v}}{Dt} = \bm{0},
\label{equ:momentum_conservation}
\end{equation}
where we ignore the inertia of fluids for the overdamped condition.
$\bm{\sigma}=-p\bm{I}+\bm{\tau}$ is the total stress, where $\bm{I}$ is identity tensor. 
$\rho$ and $\bm{g}$ are the mass density and gravity acceleration, respectively.
Due to the absence of $x$ dependence, the equation of motion \eqref{equ:momentum_conservation} is simplified into
\begin{subequations}\label{equ:equation_of_motion}
    \begin{align}
        &\text{$x$ direction:}\,
        \frac{\partial \tau_{yx}}{\partial y}+\frac{\partial \tau_{zx}}{\partial z}+\rho g\sin(\theta)=0,
        \label{equ:eom_x}\\
        &\text{$y$ direction:}\,
        -\frac{\partial p}{\partial y}+\frac{\partial \tau_{yy}}{\partial y}+\frac{\partial \tau_{zy}}{\partial z}=0, \label{equ:eom_y}\\
        &\text{$z$ direction:}\,
        -\frac{\partial p}{\partial z}+\frac{\partial \tau_{yz}}{\partial y}+\frac{\partial \tau_{zz}}{\partial z}-\rho g\cos(\theta)=0,\label{equ:eom_z}
    \end{align}
\end{subequations}
with the boundary condition by assuming that there is no velocity gradient on the free surface
\begin{equation}\label{equ:boundary_condition}
 \tau_{ij}|_{z=h}=0,\quad p|_{z=h}=p_\text{atm},
\end{equation}
where $i,j=x,y,z$ and $p_\text{atm}$ is the atmospheric pressure.


One feature of the inclined flow in initial states is that the thickness of fluids is much smaller than a characteristic length scale (defined by a wavelength of typical interfacial disturbances along $y$ direction). 
Here, we introduce a characteristic thickness scale $\mathcal{H}$ and a characteristic length scale $\mathcal{L}$ along $y$.
The discrepancy of scales allows one to introduce a small parameter as the aspect ratio $\epsilon=\mathcal{H}/\mathcal{L} \ll 1$.
Therefore, the lubrication analysis\,\cite{RevModPhys.81.1131} is valid. 
Using the two length scales, the dimensionless quantities below can be defined by
\begin{equation}\label{equ:nondim_lengths}
 \tilde{z}=\frac{z}{\mathcal{H}},
 \quad
 \tilde{h}=\frac{h}{\mathcal{H}},
 \quad
 \tilde{y}=\frac{y}{\mathcal{L}}=\epsilon\frac{y}{\mathcal{H}}.
\end{equation}
Meanwhile, a stress unit is introduced $\mathcal{S}=\rho g \mathcal{H} \sin (\theta )$. 
The dimensionless stress
can be defined by
\begin{equation}
 \tilde{\tau}_{ij}=\frac{\tau_{ij}}{\mathcal{S}},
 \quad
 \tilde{p}=\epsilon\frac{p}{\mathcal{S}}.
\end{equation}

Substituting these dimensionless quantities into eq.\,\eqref{equ:equation_of_motion}, the equation becomes a dimensionless form with coefficients composed by the small parameter $\epsilon$.
Therefore, the solution of the equation must be a function of the small parameter $\epsilon$. 
Hence, the solution of stress and pressure can then be expanded into a Taylor series. 
Here, we just keep the terms up to the first order for a linear instability analysis 
\begin{subequations}\label{equ:stress_pressure_expand}
\begin{gather}
\tilde{\tau}_{ij}=\tilde{\tau}_{ij}^{(0)}+\epsilon\tilde{\tau}_{ij}^{(1)}+o(\epsilon), \label{equ:e10aa}\\
\tilde{p}=\tilde{p}^{(0)}+\epsilon\tilde{p}^{(1)}+o(\epsilon).
\end{gather}
\end{subequations}
By using eqs.~\eqref{equ:nondim_lengths}--\eqref{equ:stress_pressure_expand}, 
\eqref{equ:equation_of_motion} becomes
\begin{widetext}
\begin{subequations}\label{equ:equation_of_motion_perturb}
    \begin{align}
        &\text{$x$ direction:}\quad
        \left(\epsilon \frac{\partial \tilde{\tau }_{yx}^{(0)}}{\partial \tilde{y}}+\epsilon ^2\frac{\partial \tilde{\tau }_{yx}^{(1)}}{\partial \tilde{y}}\right)+\left(\frac{\partial \tilde{\tau }_{zx}^{(0)}}{\partial \tilde{z}}+\epsilon\frac{\partial \tilde{\tau }_{zx}^{(1)}}{\partial \tilde{z}}\right)+ 1=0, \\
        &\text{$y$ direction:}\quad
        \left(-\frac{\partial \tilde{p}^{(0)}}{\partial \tilde{y}}-\epsilon\frac{\partial \tilde{p}^{(1)}}{\partial \tilde{y}}\right)+\left(\epsilon\frac{\partial \tilde{\tau }_{yy}^{(0)}}{\partial \tilde{y}}+\epsilon^2\frac{\partial \tilde{\tau }_{yy}^{(1)}}{\partial \tilde{y}}\right)+\left(\frac{\partial \tilde{\tau }_{zy}^{(0)}}{\partial \tilde{z}}+\epsilon\frac{\partial \tilde{\tau }_{zy}^{(1)}}{\partial \tilde{z}}\right)=0, \\
        &\text{$z$ direction:}\quad
        \left(-\frac{\partial \tilde{p}^{(0)}}{\partial \tilde{z}}-\epsilon\frac{\partial \tilde{p}^{(1)}}{\partial \tilde{z}}\right)+\left(\epsilon^2\frac{\partial \tilde{\tau }_{yz}^{(0)}}{\partial \tilde{y}}+\epsilon^3\frac{\partial \tilde{\tau }_{yz}^{(1)}}{\partial \tilde{y}}\right)+\left(\epsilon\frac{\partial \tilde{\tau }_{zz}^{(0)}}{\partial \tilde{z}}+\epsilon^2\frac{\partial \tilde{\tau }_{zz}^{(1)}}{\partial \tilde{z}}\right)- \epsilon\frac{1}{\tan(\theta)}=0,
    \end{align}
\end{subequations}
\end{widetext}
and the boundary condition \eqref{equ:boundary_condition} becomes
\begin{subequations}\label{equ:boundary_condition_perturb}
\begin{align}
&\tilde{\tau}_{ij}^{(0)}|_{\tilde{z}
=\tilde{h}}+\epsilon\tilde{\tau}_{ij}^{(1)}|_{\tilde{z}=\tilde{h}}=0, \\
&\tilde{p}^{(0)}|_{\tilde{z}=\tilde{h}}+\epsilon\tilde{p}^{(1)}|_{\tilde{z}
=\tilde{h}}=\tilde{p}_\text{atm}.
\end{align}
\end{subequations}

The terms in series of $\epsilon^n$ ($n=0, 1, \dotsc$) are linearly independent, 
which means the coefficient of $\epsilon^n$ in each term should be zero. 
Thus, eq.\,\eqref{equ:equation_of_motion_perturb} and the boundary condition~\eqref{equ:boundary_condition_perturb} are decomposed into a series of recursion equations (details in Appendix \ref{app}). 
Therefore, using eq.\,\eqref{equ:stress_pressure_expand}, we can obtain the first-order approximation solution from the perturbation results eqs.~\eqref{equ:e15} and \eqref{equ:e18},
\begin{subequations}
    \begin{align}
        &\tilde{\tau}_{zx} = \tilde{h}-\tilde{z}+\epsilon\int _{\tilde{z}}^{\tilde{h}}\frac{\partial \tilde{\tau }_{yx}^{(0)}}{\partial \tilde{y}}d\tilde{z} +o(\epsilon),\\
        &\tilde{\tau}_{zy} = -\epsilon\left(\frac{\tilde{h}-\tilde{z} }{\tan (\theta )}\frac{\partial \tilde{h}}{\partial \tilde{y}}+\int _{\tilde{z}}^{\tilde{h}}\frac{\partial \tilde{N}_2^{(0)}}{\partial \tilde{y}}d\tilde{z}\right)+o(\epsilon), \label{equ:e19b} \\
        &\tilde{p} = \tilde{p}_\text{atm}+\epsilon\left(\frac{\tilde{h}-\tilde{z} }{\tan (\theta )}+\tilde{\tau }_{zz}^{(0)}\right)+o(\epsilon).
    \end{align}
\end{subequations}

\subsection{The constitutive relation and its evidences}
\label{subsec: constitutive relation}

\begin{figure*}[htpb]
\centering
\includegraphics[width=0.5\textwidth]{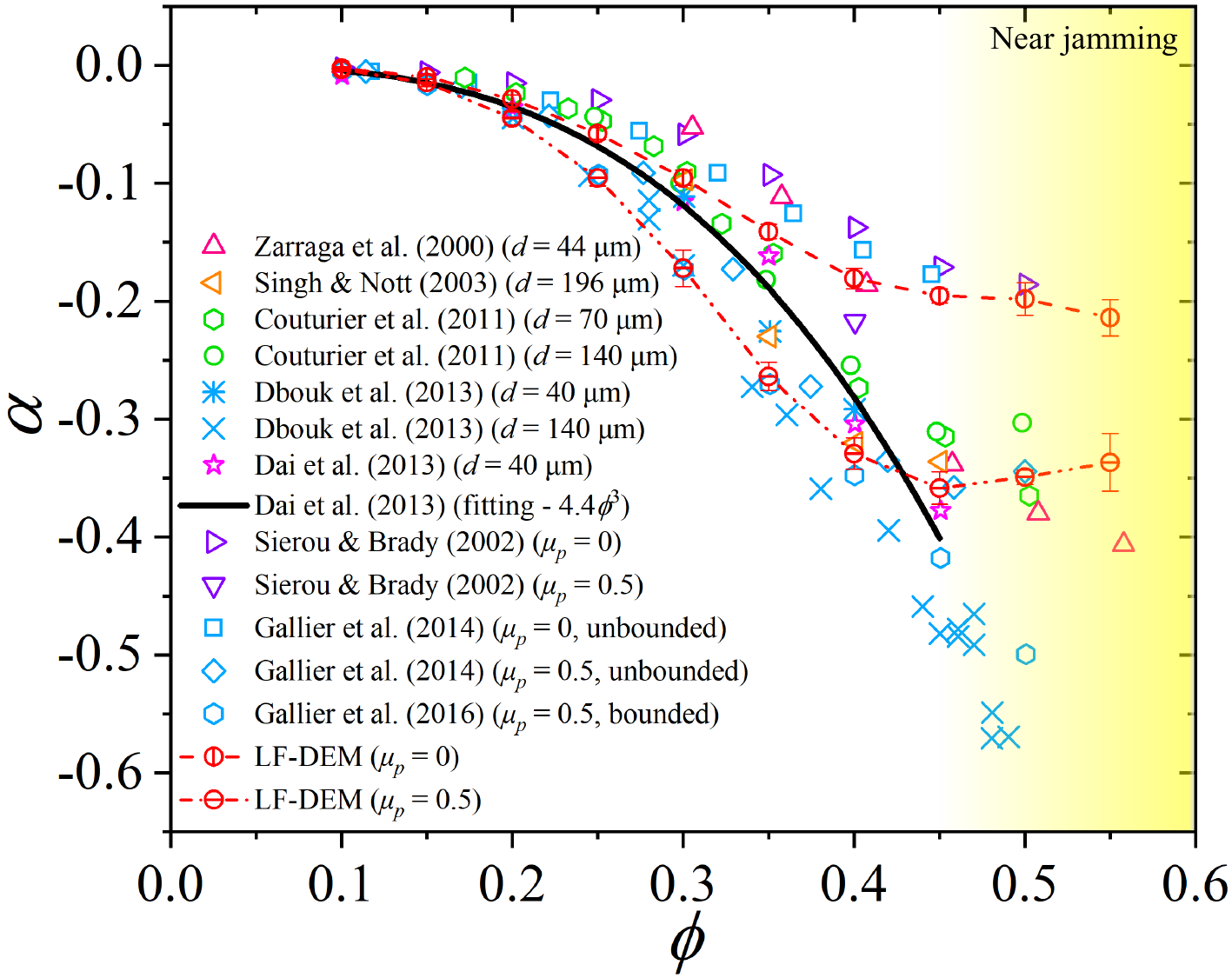}
\caption{\label{fig:f3} 
The second normal-stress-difference ratio $\alpha \coloneqq N_2 / \tau_{zx}$ versus volume fraction $\phi$. 
Experimental data from Zarraga~\textit{et al.}\,(2000)\cite{10.1122/1.551083}
with glass spheres of diameter $d = \SI{44}{\micro\metre}$, 
Singh \& Nott\,(2003)\cite{singh2003experimental} 
with PMMA spheres of diameter $d = \SI{196}{\micro\metre}$, 
Couturier~\textit{et al.}\,(2011)\cite{couturier2011suspensions} with PS spheres of diameters $d = 70$ and $\SI{140}{\micro\metre}$, 
Dbouk~\textit{et al.}\,(2013)\cite{dbouk2013normal} with PS spheres of diameters $d = 40$ and $\SI{140}{\micro\metre}$,
and Dai~\textit{et al.}\,(2013)\cite{dai2013viscometric} with PS spheres of diameter $d = \SI{40}{\micro\metre}$.
The fitting in \citet{dai2013viscometric}\,(2013) for the range $0.1\leq\phi\leq0.45$ 
is also shown by the solid line.
Numerical simulations from Sierou \& Brady~(2002)\cite{sierou2002rheology}, Gallier~\textit{et al.}\,(2014)\cite{gallier2014rheology} without and with friction ($\mu_{\mathrm{p}} = 0$ and $0.5$),
and Gallier~\textit{et al.}\,(2016)\cite{Gallier_2016} under a confinement (bounded) 
with friction ($\mu_{\mathrm{p}} = 0.5)$.
%
We also perform the LF-DEM simulation\cite{Romain2015Discontinuous,seto2018normal} to evaluate $\alpha$ without and with the particle friction ($\mu_{\mathrm{p}} = 0$ and $0.5$) (dashed lines).
 }
\end{figure*}

As discussed in section \ref{sec: intro}, the second normal stress difference of suspensions is coupled to shear stress with a linear relationship in a simple shear flow, i.e.,
\begin{equation}\label{equ:N2}
    N_2=\alpha(\phi)\tau_{zx},
\end{equation}
where $N_2 \coloneqq \tau_{zz}-\tau_{yy}$ is the second normal stress difference and $\tau_{zx}$ is the shear stress, since the `flow,' `gradient,' and `vorticity' directions are $x$, $z$, and $y$ directions in this work when the free surface is near the initial flat configuration. 
%


Suppose that some rate-dependent property of suspensions, such as shear-thickening or shear-thinning, is given as $\tau_{zx}=\eta(\tau_{zx})\dot{\gamma}$.
One critical underlying assumption of eq.\,\eqref{equ:N2} is that the $N_2$ has also the similar rate-dependent property $N_2 \propto \eta(\tau_{zx})\dot{\gamma}$. 
If such a condition is satisfied, 
one can introduce a rate-independent
material parameter $\alpha$. 
Otherwise, $\alpha$ becomes rate-dependent. 
Thus, in this work, we focus on the rate-independent $\alpha$, i.e., a class of suspension where shear stress and anisotropic normal stresses are linked to each other to hold \eqref{equ:N2}.
Such a class of suspension has been widely investigated in both experiments and simulations. 
%


For Newtonian fluids in the dilute limit $\phi \to 0$, one has $\alpha=0$ because of the vanished $N_2$.
Thus, the $\phi$ dependence of $\alpha$ is expected because there must be a transition from the Newtonian fluids to the dense suspension as $\phi$ increases. 
More detailed investigations show that $\alpha$ is a negative value and solely dependent on $\phi$\,\cite{dai2013viscometric, guazzelli2018rheology, cwalina2014material}.
Fig.\,\ref{fig:f3} shows experimental and simulation data collected from various references\,\cite{guazzelli2018rheology}.
In the dilute situation, the $\phi$ dependence of $\alpha$ follows the quadratic relation since the pairwise interactions are dominated\,\cite{wilson2005analytic}.
There are some empirical fitting results to illustrate the $\phi$ dependence of $\alpha$, such as $\alpha = -4.4\phi^3$ in the range $0.1 \leq \phi \leq 0.45$ by \citet{dai2013viscometric}(shown by the solid line in Fig.\,\ref{fig:f3}).
We also confirm the basic tendency by using the LF-DEM simulation\,\cite{Romain2015Discontinuous,seto2018normal} without and with particle friction ($\mu_{\mathrm{p}} = 0$ and $0.5$), as shown by the dashed lines in Fig.\,\ref{fig:f3}.
The averages and standard deviations (error bars) are taken over 10 independent simulations for each condition (further simulation details are given here\,%
\footnote{In our LF-DEM simulation, the lubrication outer cutoff is $0.5 a$ ($a$ particle radius), and the divergence cutoff is $10^{-3}a$. 
The particle-overlap spring constant is $k_{\mathrm{n}} = 50000k_0$, where $k_0 \equiv 6 \pi \eta_0 a \dot{\gamma} $.
The maximum particle displacement at each time step is set to $10^{-4}a$. 
The Einstein term $2.5\eta_0 \phi$ is added for the viscosity.}).
%
%



Therefore, the constitutive relation eq.\,\eqref{equ:N2} with the informative connection to $\phi$ is valid for further analysis near the initial flat configuration of the free surface. 
By using eq.\,\eqref{equ:e10aa} and the solution~\eqref{equ:e15a}, we have
\begin{equation}\label{eq:15.1}
\tilde{N}_2^{(0)}=\alpha(\phi)\tilde{\tau}_{zx}^{(0)}
=\alpha(\phi)\big(\tilde{h}-\tilde{z}\big).
\end{equation}
Inserting eq.\,\eqref{eq:15.1} into \eqref{equ:e19b}, the shear stress is written as
\begin{equation}
	\tilde{\tau}_{zy} = -\epsilon\left(\frac{1 }{\tan (\theta )}+\alpha(\phi)\right)\big(\tilde{h}-\tilde{z}\big)\frac{\partial \tilde{h}}{\partial \tilde{y}} + o(\epsilon).
\end{equation}
Therefore, the first-order approximation of the shear rate along the $y$ direction is
\begin{align}\label{equ:e24}
 \frac{\partial \tilde{v}_y}{\partial \tilde{z}}&=\frac{\tilde{\tau}_{zy}}{\tilde{\eta}(\tilde{\tau}_{zy})}-\epsilon^2\frac{\partial \tilde{v}_z}{\partial \tilde{y}}\approx\frac{\tilde{\tau}_{zy}}{\tilde{\eta}(0)} \notag\\
    &\approx-\frac{\epsilon}{\tilde{\eta}(0)}\left(\frac{1 }{\tan (\theta )}+\alpha(\phi)\right)
    \big(\tilde{h}-\tilde{z}\big)
    \frac{\partial \tilde{h}}{\partial \tilde{y}},
\end{align}
where we have ignored the stress dependence of viscosity since $\tilde{\tau}_{zy}$ is essentially small in initial state.

\section{The origin of the instability}\label{sec:s3}

Integrating both sides of eq.\,\eqref{equ:e24} from $\tilde{z}=0$ to $\tilde{z}$ ($\leq\tilde{h}$) and using the non-slip condition $\tilde{v}_y|_{\tilde{z}=0}=0$, one has
\begin{equation}
\tilde{v}_y=-\frac{\epsilon}{\tilde{\eta}(0)}
 \left(\frac{1 }{\tan (\theta )}+\alpha(\phi)\right)
 \left(\tilde{h}\tilde{z}-\frac{\tilde{z}^2}{2}\right)
 \frac{\partial \tilde{h}}{\partial \tilde{y}}.
\end{equation}
Combined with the continuity equation~\eqref{equ:free_surface_continuity} in a dimensionless form, one has
\begin{align}
 0 &= \frac{\partial \tilde{h}}{\partial \tilde{t}}
     +\frac{\partial}{\partial \tilde{y}}
    \int_0^{\tilde{h}}\tilde{v}_yd\tilde{z} \notag\\
   &=\frac{\partial \tilde{h}}{\partial \tilde{t}}
   -\frac{\epsilon}{3\tilde{\eta}(0)}\left(\frac{1 }{\tan (\theta )}+\alpha(\phi)\right)
   \frac{\partial}{\partial \tilde{y}}
   \biggl(\tilde{h}^3\frac{\partial \tilde{h}}{\partial \tilde{y}}\biggr).
   \label{equ:e26}
\end{align}

At the very beginning, the flat free surface starts to be distorted due to the fluctuation, and the instability makes the distortion keep growing.
Since the fluctuation is small ($\Tilde{h}'\ll 1$), one has
\begin{equation}
 \Tilde{h}=1+\Tilde{h}'.
\end{equation}
Eq.\,\eqref{equ:e26} then becomes
\begin{equation}\label{equ:e28}
 \frac{\partial \Tilde{h}'}{\partial \tilde{t}}=\frac{\epsilon}{3\tilde{\eta}(0)}\left(\frac{1 }{\tan (\theta )}+\alpha(\phi)\right)\frac{\partial^2 \Tilde{h}'}{\partial \tilde{y}^2}.
\end{equation}


Eq.\,\eqref{equ:e28} is a diffusion-like equation for a fluctuation $\Tilde{h}'$. 
If the `diffusion coefficient' is greater than zero, the fluctuation will decay with time, indicating that the system is stable for some perturbations. 
However, it becomes unstable with a negative `diffusion coefficient,' which leads to an instability condition
\begin{equation}
 \frac{1}{\tan (\theta )}+\alpha(\phi)<0.
\end{equation}
Therefore, the flow becomes unstable along $y$ direction if the inclined angle is larger than the critical angle $\theta_{\mathrm{c}}$:
\begin{equation} 
\label{criterion} \theta>\theta_{\mathrm{c}}=\arctan\biggl(-\frac{1}{\alpha(\phi)}\biggr).
\end{equation}

This is a different criterion of instability from that of the 
roll wave\,\cite{darbois2020surface} based on the 
key assumption\,\eqref{equ:N2}.
The present results from both the experiment and the linear stability analysis indicate that the $N_2$ mechanism can be safely disentangled from the DST properties of suspensions.
The rate dependence (shear thickening or shear thinning) is not a necessary condition in the linear instability analysis if the material parameter $\alpha$ is rate independent.


Thus, the theory predicts an instability caused by a negative value of $\alpha $ in \eqref{equ:N2}, which can be considered as the concept called dilatancy\,\cite{Reynolds_1885}; 
impenetrable solid particles under shear tend to climb on contacting particles and dilate along the gradient direction of the flow.
%
%
The dilating stress lifts some thick parts of the fluid film to form a ridge, but the rest of the parts need to subside due to the incompressibility of the fluid. 
Such a height profile can develop with a positive feedback manner if the gravity component perpendicular to the wall is weak enough.
%
%
%

For Newtonian fluids ($\phi \rightarrow 0$), $\alpha=0$ is expected, and the fluids will never appear in the ridge formation because the angle $\theta > \ang{90}$ is required according to eq.~\eqref{criterion}.
In our experiment, the sample with $\phi\approx 0.37$ is prepared to show the ridge-like formation. 
Based on the fitting correlation in Fig.\,\ref{fig:f3}, one can obtain $\alpha \approx -0.22$ and $\theta_{\mathrm{c}} \approx \ang{77.4}$, which agree with 
the experimental estimation in section \ref{sec: experiment}.

The macroscopic relation of the $\phi$ dependence of $\alpha$ could be complicated since the fitting cubic relation is not for both dilute and dense limits.
The valid $\alpha(\phi)$ will be helpful to explain the phase diagram of the rheological morphogenesis in parameter space of $\theta$-$\phi$.
Combining the two instability criteria for the two simplest horizontal and vertical wave patterns, the more complex pattern formation may be understood by triggering the two instabilities under different $\theta$ and $\phi$ conditions.

It is also worth mentioning 
that we rely on uncontrolled natural fluctuations for the initial distortion
and visual check for the amplitude growth,
which tends to overestimate the critical angle.
For more accurate determination, 
it is desirable to develop experimental setups to control perturbation and to monitor the profile, like the ones by \citet{darbois2020surface}.
%
%
%
In this work, we stayed concision to show essential elements that can cause ridge instability.
We need to have more systematic experimental investigations to explore the parameter space in the future.

\section{Conclusion}

In this work, we observed a novel instability where a particle suspension flowing down an inclined plane formed ridges parallel to the flow direction. We propose that the ridges form due to a dilation effect caused by the second normal stress difference $N_2$.
We obtained two force balance equations for shear stresses $\tau_{zx}$ and $\tau_{zy}$ from the equation of motion based on lubrication analysis without inertia. 
The linear relation between the second normal stress difference $N_2$ and shear stress $\tau_{zx}$ is used as a key assumption. 
Based on these relations, we can obtain a different instability criterion from that of 
the roll wave\,\cite{darbois2020surface} by linear instability analysis. 
Therefore, the ratio between $N_2$ and $\tau_{zx}$ is a key parameter to control the instability that distorts the initial flat configuration along the perpendicular direction of flow and determines the critical angle $\theta_{\mathrm{c}}$. 
Further experiments with higher precision are needed to confirm whether the instability we have shown is the reason for the observed ridge formation.

\begin{acknowledgments}
We thank Dan Curtis for the discussions and assistance with rheometry measurements
and Yang Cui for the discussions and for providing important literature.
%
We wish to acknowledge the support of the Natural Science Foundation of China (NSFC) (Nos.\,12247174, 12150610463, and 12174390) and Wenzhou Institute, University of Chinese Academy of Sciences (WIUCASQD2020002).
Financial support through the BERC 2022-2025 program and by the Spanish State Research Agency through BCAM Severo Ochoa Excellence Accreditation CEX2021-001142-S/MICIN/AEI/10.13039/501100011033 and through the project PID2020-117080RB-C55 (`Microscopic foundations of soft-matter experiments: computational nano-hydrodynamics') funded by AEI--MICIN and acronym `Compu-Nano-Hydro' are also gratefully acknowledged. 
PA and BS acknowledge funding from the Engineering and Physical Sciences Research Council [EP/S034587/1].

\end{acknowledgments}

\appendix

\section{Viscosity of the used cornstarch suspensions} \label{viscos}

\begin{figure}[htpb]
\centering
\includegraphics[width=0.45\textwidth]{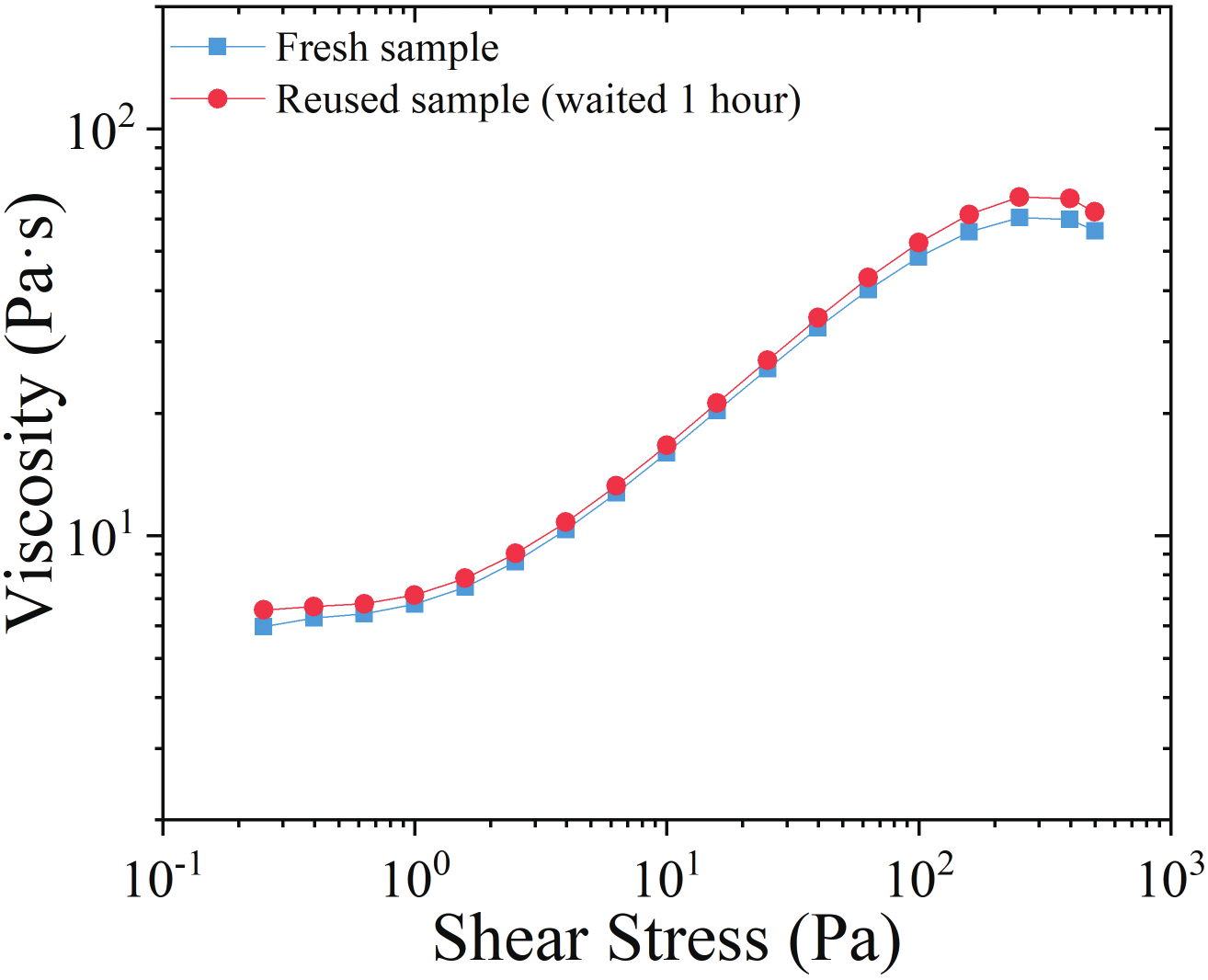}
\caption{\label{fig:f4} 
The viscosity of the used cornstarch suspension was measured (Fresh sample) and measured again (Reused sample) waited another 1 hour.}
\end{figure}

The viscosity of the sample has been measured twice with a one-hour time interval (to confirm no significant aging effect during slope flow experiments) for sure, as shown in Fig.\,\ref{fig:f4}, which appears to be CST. 
The inclined flow is a stress-controlled flow driven by gravity. 
Thus, stress-controlled experiments have been conducted to measure the viscosity on ARES-G2 equipped with parallel plates ($\SI{60}{mm}$ diameter and $\SI{1}{mm}$ gap) at $\SI{25}{\degreeCelsius}$.

\section{Perturbation Equations}\label{app}

The vanished coefficient of $\epsilon^n$ in eqs.~\eqref{equ:equation_of_motion_perturb} 
and \eqref{equ:boundary_condition_perturb} 
gives the relations by each order as follows:

\noindent
(1) For term with $\epsilon^0$
\begin{subequations}
    \begin{align}
        &\text{$x$ direction:}\quad
        \frac{\partial \tilde{\tau }_{zx}^{(0)}}{\partial \tilde{z}}+1=0, \\
        &\text{$y$ direction:}\quad
        -\frac{\partial \tilde{p}^{(0)}}{\partial \tilde{y}}+\frac{\partial \tilde{\tau }_{zy}^{(0)}}{\partial \tilde{z}}=0, \\
        &\text{$z$ direction:}\quad
        -\frac{\partial \tilde{p}^{(0)}}{\partial \tilde{z}}=0,
    \end{align}
\end{subequations}
with boundary condition
\begin{equation}
\tilde{\tau}_{ij}^{(0)}|_{\tilde{z}
=\tilde{h}}=0,\quad \tilde{p}^{(0)}|_{\tilde{z}
=\tilde{h}}=\tilde{p}_\text{atm}.
\end{equation}
The solution of the equations is
\begin{subequations}\label{equ:e15}
   \begin{align}
    \tilde{\tau }_{zx}^{(0)}&=\tilde{h}-\tilde{z}, \label{equ:e15a}\\
    \tilde{\tau }_{zy}^{(0)}&=0, \\
    \tilde{p}^{(0)}&=\tilde{p}_\text{atm}.
    \end{align}
\end{subequations}

\noindent
(2) For term with $\epsilon^1$
\begin{subequations}
    \begin{align}
        &\text{$x$ direction:}\quad
        \frac{\partial \tilde{\tau }_{yx}^{(0)}}{\partial \tilde{y}}+\frac{\partial \tilde{\tau }_{zx}^{(1)}}{\partial \tilde{z}}=0, \\
        &\text{$y$ direction:}\quad
        -\frac{\partial \tilde{p}^{(1)}}{\partial \tilde{y}}+\frac{\partial \tilde{\tau }_{yy}^{(0)}}{\partial \tilde{y}}+\frac{\partial \tilde{\tau }_{zy}^{(1)}}{\partial \tilde{z}}=0, \\
        &\text{$z$ direction:}\quad
        -\frac{\partial \tilde{p}^{(1)}}{\partial \tilde{z}}+\frac{\partial \tilde{\tau }_{zz}^{(0)}}{\partial \tilde{z}}-\frac{1 }{\tan (\theta )}=0,
    \end{align}
\end{subequations}
with boundary conditions,
\begin{equation}
\tilde{\tau}_{ij}^{(1)}|_{\tilde{z}=\tilde{h}}=0,
\quad \tilde{p}^{(1)}|_{\tilde{z}=\tilde{h}}=0.
\end{equation}
The solution of the equations is
\begin{subequations}\label{equ:e18}
    \begin{align}
        \tilde{\tau }_{zx}^{(1)}&=\int _{\tilde{z}}^{\tilde{h}}\frac{\partial \tilde{\tau }_{yx}^{(0)}}{\partial \tilde{y}}d\tilde{z}, \\
        \tilde{\tau }_{zy}^{(1)}&=-\frac{\tilde{h}-\tilde{z} }{\tan (\theta )}\frac{\partial \tilde{h}}{\partial \tilde{y}}-\int _{\tilde{z}}^{\tilde{h}}\frac{\partial \tilde{N}_2^{(0)}}{\partial \tilde{y}}d\tilde{z}, \\
        \tilde{p}^{(1)}&=\frac{\tilde{h}-\tilde{z} }{\tan (\theta )}+\tilde{\tau }_{zz}^{(0)},
    \end{align}
\end{subequations}
where the zero-th order of second normal stress difference is $\tilde{N}_2^{(0)} \coloneqq \tilde{\tau}_{zz}^{(0)}-\tilde{\tau}_{yy}^{(0)}$ since the conventional definition is the stress difference between `gradient' and `vorticity' directions.

\nocite{*}
\bibliography{reference}

\end{document}